\documentclass[aps,prd,preprint,groupedaddress,showpacs,nofootinbib]{revtex4-1}                 % ... or a4paper or a5paper or ... 
\usepackage{natbib}
\usepackage{graphicx}% Include figure files
\usepackage{dcolumn}% Align table columns on decimal point
\usepackage{bm}% bold math
\usepackage{xcolor}
\usepackage[utf8]{inputenc}
\usepackage[T1]{fontenc}
\usepackage{float}
\usepackage{amsmath}
\usepackage{hyperref}% add hypertext capabilities
\begin{document}
\title{Screening corrections to Electron Capture Rates and resulting  constraints on Primordial Magnetic Fields}

\author{Yudong Luo$^{1,2}$}
\email{ydong.luo@nao.ac.jp}
\author{Michael A. Famiano$^{1,3}$}
\email{michael.famiano@wmich.edu}
\author{Toshitaka Kajino$^{1,2,4}$}
\email{kajino@nao.ac.jp}
\author{Motohiko Kusakabe$^{1,4}$}
\email{kusakabe@buaa.edu.cn}
\author{A. Baha Balantekin$^{1,5}$}%
\email{baha@physics.wisc.edu}

\affiliation{$^1$National Astronomical Observatory of Japan,
2-21-1 Osawa, Mitaka, Tokyo 181-8588, Japan}
\affiliation{$^2$Graduate School of Science,
The University of Tokyo, 7-3-1 Hongo, Bunkyo-ku, Tokyo 113-0033, Japan}
\affiliation{$^3$Department of Physics, Western Michigan University, 
Kalamazoo, Michigan 49008,USA}
\affiliation{$^4$School of Physics, and International Research Center for Big-Bang Cosmology and Element Genesis, Beihang University, 37 Xueyuan Rd., Haidian-district, Beijing 100083 China}
\affiliation{$^5$Department of Physics, University of Wisconsin, Madison, WI 53706 USA
}
\date{\today}
\begin{abstract}%
We explore screening effects arising from a relativistic magnetized plasma with applications to Big Bang nucleosynthesis (BBN). %Specifically, due to their small magnetic moments, energies of electrons and positrons can be easily quantized via Landau quantization. 
The screening potential which depends on the thermodynamics of charged particles in the plasma is altered by the magnetic field. We focus on the impact of screening on the electron capture interaction. Taking into account the correction in BBN arising from a homogeneous primordial magnetic field (PMF), we constrain 
the epoch at which the PMF was generated and its strength during BBN. Considering such screening corrections to the electron capture rates and using up-to-date observations of primordial elemental abundances, we also discuss the possibility of solving the problem of under-estimation of the deuterium abundance. We find for certain values of the PMF strength predicted D and $^4$He abundances are both consistent with the observational constraints.
 \end{abstract}
\maketitle
\section{Introduction}
Big Bang nucleosynthesis (BBN) is one of the three key pieces of evidence of the hot big bang model, providing a robust tool in order to probe the physics of the early Universe. Theoretical calculations of light element abundances (namely, D, $^4$He and $^7$Li) in the standard BBN (SBBN) model are well-characterized  \cite{Cyburt:2016cr,2017IJMPE..2641001M,Pitrou2019}. Although there is a long-standing problem with the SBBN prediction of the primordial $^7$Li abundance, which is 4 times higher than the observations \cite{1982A&A...115..357S, 2010A&A...522A..26S,Field2011}, the deuterium and $^4$He observations now have reached an accuracy on the order of a percent, consistent with theoretical predictions. 

The SBBN model has three parameters: the effective neutrino number, $N_\nu$, the neutron lifetime, $\tau_n$, and the 
baryon to photon ratio (the baryonic density) of the Universe, $\eta$. All these three parameters have been fairly well
determined from experiments \cite{Patrignani,Descov2004,Coc2015} and the analysis of cosmic microwave background (CMB)
power spectrum analysis \cite{2013ApJS..208...20B,2016A&A...594A..13P}. Therefore, the study of the thermonuclear
reaction rates becomes significant. For deuterium, Coc et al. (2015)~\cite{Coc2015} re-evaluated the uncertainties of D
production cross sections and obtained even smaller uncertainties in the D/H predictions. The theoretical uncertainty
of the predicted of abundance
of $^4$He mainly arises from the uncertainty in the  experimental value of the neutron lifetime \cite{CPC2014} which
includes the numerous corrections of the theoretical weak–interaction rates. Pitrou et al. (2019)
\cite{Pitrou2019} have investigated such corrections and successfully reached a precision of better than 0.1\%.

Recently, charge screening from both ionized nuclei and electrons in relativistic electron-positron plasmas have been
discussed and applied to  the determination of thermonuclear reaction rates
\cite{2011PhRvC..83a8801W,2016PhRvC..93d5804F}. This effect turned out to be negligible during the BBN epoch because the
plasma is in a high temperature, low density state, and the distance between electrons or positrons and nuclei is so 
large that the screening effect on the Coulomb potential is not significant. However, at the epoch before weak 
decoupling, i.e., $t\lesssim1 \rm sec$ and $kT>1\ \rm MeV$, the density is much higher compared with the later BBN 
epoch, and there is also a large number of electrons and positrons. The screening effect in a relativistic 
electron-positron plasma could affect weak interaction rates by changing the electron and positron energy distributions.
Such screening corrections to the electron capture rates have been studied and applied to stellar nucleosynthesis
\cite{Itoh}; however, this approach is not suitable for the relativistic non-degenerate electron-positron plasma.

On the other hand, non-standard BBN models including extra physics such as the primordial magnetic field (PMF) 
\cite{OConnell1969,Greenstein1969,Matese1970,Cheng:1993kz,1995APh.....3...95G,1996PhRvD..54.7207K,Grasso:1996kk,Cheng:1996yi,1997PhRvD..56.3766K,Grasso:2000wj,Kawasaki:2012kn,Yamazaki2012} have ever been studied, and a possible moderation
of the cosmic lithium problem has been investigated \cite{Luo}. Under a strong magnetic field, the weak reaction rates are affected since the charged-particle distribution functions are altered \cite{OConnell1969,1996PhRvD..54.7207K,Cheng:1996yi}. In addition, the energy density of the field affects the cosmic expansion rate \cite{Greenstein1969,Matese1970}, which has been considered as one of the most important effects of the PMF on BBN. The effects of the PMF on the cosmic expansion rate and temperature evolution during BBN come through the change in the momentum distribution function of electrons and positrons \cite{Grasso:1996kk,Grasso:2000wj}. A full formulation of the PMF effects on the cosmic expansion rate and the temperature evolution has been derived \cite{Kawasaki:2012kn}, which shows primordial abundances of all light nuclei, i.e., D, $^{3,4}$He and $^{6,7}$Li, as a function of the PMF amplitude derived from a consistent numerical calculation taking into account changes in evolution of the electron chemical potential and the baryon-to-photon ratio induced by the PMF.

 It has been found that a $\mu$G scale magnetic field exists in the Galaxy  via the both Faraday rotation (\cite{Davies1968rm} and references therein) and Zeeman effect \cite{Verschuur1968,Davies1968z}. Observations of intermediate and high redshift galaxies also indicate the existence of such large magnetic fields. The PMF is considered to be a seed of these Galactic magnetic fields which have amplified via the dynamo mechanism (see e.g., \cite{Kronberg:1993vk,Grasso:2000wj,Widrow:2011hs} for reviews). A rather large PMF is needed as a seed since only a short duration time is available for the amplification of the magnetic field \cite{Kronberg:1993vk,Grasso:2000wj,Widrow:2011hs} from formations of the observed galaxies until the epochs corresponding to their redshifts. The PMF is thought to be generated from cosmological inflation, phase transitions and/or astrophysical processes \citep{Dolgov2001,PhysRevLett.95.121301,Ichiki827,Durrer:2013ec,2016RPPh...79g6901S,2016PhRvD..93d3004Y}. Once the seed of magnetic field is generated, it is possible later on to be amplified via magneto-hydrodynamic (MHD) processes \cite{2005PhR...417....1B,1997PhLB..390...87D}. Although the damping of the PMF was also studied previously \cite{1998PhRvD..57.3264J}, this PMF is considered to be the seed of the Galactic magnetic field which was amplified via the dynamo mechanism. After the epoch of photon last scattering ($z\sim1100$), the CMB power spectrum provides us with an observable constraint on the energy density of the PMF \cite{2008PhRvD..77d3005Y,2010PhRvD..81d,2012PhRe}. Meanwhile, the primordial elemental abundances also can constrain the PMF strongly.

The PMF is studied at two epochs in the cosmic evolutionary history. The first is the PMF generated during the inflation and phase transition epoch. Since the horizon during the inflation and phase transition is much smaller than the typical length scale of the present-day PMF observation, a super-horizon PMF generated during inflation and (or) phase transition \citep{1988PhRvD..37.2743T,1993PhRvD..48.2499D,2009JCAP...08..025D} is possible. This kind of magnetic field is "frozen-in" with the dominant fluids. A PMF on super-horizon scales during BBN effectively works as a homogeneous field. The other PMF is the inhomogeneous PMF generated at later epochs. In one model proposed by Dolgov and Grasso~\citep{Dolgov2001}, the smaller scale of the PMF fluctuations inside the co-moving horizon is expected to survive during the BBN epoch due to the local imbalance of lepton number. It is therefore possible to assume that the PMF energy density had an inhomogeneous distribution inside the horizon at BBN \cite{Luo}.

In either case, it is important to derive a constraint on generation epoch of the magnetic field in order to clarify the origin of
the PMF.  At higher temperature, weak interactions play a leading role in the hot relativistic plasma. Previous studies of the PMF always neglected its impact on the weak interaction \cite{Yamazaki2012,Kawasaki:2012kn} since the change of neutron fraction $X_n$ is as small as 0.01. However, the up-to-date BBN theoretical and observational constraints on primordial $^4$He abundance have reached an accuracy of $10^{-4}$, and any change in $X_n$ larger than this amount is to be constrained carefully. In this paper, we consider two aspects of the impacts made by PMF on the weak interactions. On one hand, we investigate the impact on the weak interaction from PMF directly. On the other hand, by introducing weak-interaction screening corrections, we derive weak interaction rates in the
presence of magnetic fields before and during BBN. 

This paper is organized as follows. In Section \ref{sect1}, we explain the screening effect and its correction to the electron capture reaction. We also investigate the weak interaction properties under the background PMF.  In Section \ref{sect2}, we discuss how the PMF affects the prediction of primordial light element abundances and try to provide constraints in turn on the PMF. We give the conclusion in Section \ref{sect3}.

\section{Weak Screening correction of the electron capture rate}\label{sect1}
\subsection{Weak Screening Correction}
In a hot plasma, the background charged particles can create a "screening" effect which reduces the Coulomb barrier between two fusion reactants by reducing the effective charge \citep{jancovici77,2016PhRvC..93d5804F}. The background charges include the surrounding electrons, positrons, and other nuclei. Classically, the electrostatic potential $\phi$ of a charge $ze$ in the presence of a background charge density can be computed via
the {\it Poisson-Boltzmann equation}:
\begin{equation}
\label{PB}
\nabla^2\phi(r) = -4\pi Ze^2\delta^3(\mathbf{r}) -4\pi\sum_{z\ge0} ze n_z\left[\exp\left(-\frac{ze\phi(r)}{T}\right) -1 \right]- e \left[ N(\mu + e \phi, T) - N(\mu, T) \right],
\end{equation}
where 
\begin{equation}
\label{2}
N (\mu,T) = \frac{1}{\pi^2} \int dp \> p^2 \left[ \frac{1}{e^{(E-\mu)/T} +1} - \frac{1}{e^{(E +\mu)/T}+1} \right]
\end{equation}
is the net lepton number density, 
$T$ and $\mu$ are the temperature and the chemical potential of electrons in units of MeV (hereafter, we use natural 
units, i.e. $\hbar=k=c=1$). The second term of Eq.(\ref{PB}) is a sum over all charged nuclei in the medium with charge
$ze$ and number density $n_z$. The last term includes the charge of the electrons and positrons. This is universally 
used in astrophysical calculations involving nuclear reactions. By
expanding Eq. (\ref{PB}) to lowest order in potential $\phi$, one obtains the solution as
the familiar Yukawa potential:
\begin{equation}\label{poten}
\phi(r)=\frac{Z_1Z_2e^2}{r}\exp{\left(-\frac{r}{\lambda_{TF}}\right)}.
\end{equation}
For the relativistic electron-positron plasma, the corresponding \textit{Thomas-Fermi length} can be calculated exactly
to all orders from the {\it Schwinger-Dyson equation} for the photon propagator \citep{2016PhRvC..93d5804F}. The
characteristic length scale is:
\begin{equation}\label{eq_debye}
\frac{1}{\lambda_{TF}^2}=4\pi e^2\frac{\partial}{\partial\mu}\int^\infty_0 dp\
p^2\left[\frac{1}{1+\exp{(E-\mu)/T}}-\frac{1}{1+\exp{(E+\mu)/T}}\right],
\end{equation}
where $\mu$ is the electron chemical potential. 

Screening corrections to $\beta$-decay rates have been discussed previously~\citep{Takahashi1978,Fuller1980,Liu2007}. The possible importance of the screening effects on the electron capture rates at extremely high densities
have also been investigated. However, the plasma is not degenerate \cite{Itoh1983,Itoh} in the early Universe with a high density and temperature before the completion of the $e^+ e^-$ annihilation, and non-degenerate relativistic screening corrections to the electron capture have not been well studied. In the non-degenerate environments, the distance between particles is always much smaller than $\lambda_{TF}$, therefore Eq. (\ref{poten}) can be expanded to the first order and compared with the Coulomb potential from bare nuclei.  The correction to weak screening is shown to be:
\begin{equation}\label{Correction}
\Delta V=\phi(r)-V^{bare}\approx\frac{Z_1Z_2e^2}{\lambda_{TF}}.
\end{equation}
In the early Universe, weak interactions play an important role in
calculating the proton-to-neutron ratio $n/p$. The predicted $^4$He mass fraction $Y_p$ is mainly determined by $2n/(n+p)$~\cite{Kajino2016} at the epoch of weak decoupling. When the temperature of the Universe is higher than the mass difference between proton and neutron, $q=m_p-m_n$, neutrons and protons are indistinguishable via three main weak interactions:
\begin{eqnarray}
\nonumber n+e^+\longleftrightarrow p +\bar{\nu_e},\\
\nonumber n+\nu_e\longleftrightarrow p +e^-,\\
\nonumber n\longleftrightarrow p +e^-+\bar{\nu_e}.\\
\end{eqnarray}

The cross sections for weak interactions are calculated with the V-A interaction Hamiltonian~\cite{Book}. For electron capture process, i.e., $p +e^-\to n+\nu_e$, the screening correction [Eq. (\ref{Correction})] influences the cross section through a change in the Coulomb potential. The kinetic energies of electrons around protons are shifted due to screening.  The electron capture rate on protons, $\Gamma_{pe^-\rightarrow n\nu_e}$, are:
\begin{eqnarray}\label{e_cap_noB}
\Gamma^{scr}_{pe^-\to n\nu_e}=\frac{G_F^2T_\gamma^2(g_V^2+3g_A^2)}{2\pi^3}
\int_1^{\infty} E_\nu^2\epsilon'\sqrt{\epsilon'^2-m_e^2}\ d\epsilon'\ f_{FD}(\epsilon';\mu,T_\gamma)g(E_\nu;\mu_\nu,T_\nu),
\end{eqnarray}
where $G_F$ is the {\it Fermi coupling constant}, $g_V=1.4146\times10^{-49}\rm\ erg\ cm^3$ and $g_A/g_V\sim-1.262$, $E_\nu$ is the neutrino energy, $\mu_\nu$ is the neutrino chemical potential, $T_\gamma$ and $T_\nu$ represent the photon and neutrino temperatures respectively. The notation $f_{FD}(\epsilon'; \mu, T_\gamma) =1/[{ \exp[(\epsilon' -\mu)/T_\gamma] +1}]$ is the Fermi-Dirac distribution function, 
and $g(E_\nu; \mu_\nu, T_\nu)=1-f_{FD}(E_\nu; \mu_\nu, T_\nu)$ is the Pauli blocking factor. The screening correction to the electron kinetic energy is given by $\epsilon'=\epsilon-\Delta V$. The rare three-body reaction, $pe^-\bar{\nu}_e\rightarrow n$, is ignored.

%Although the reaction $pe^-\bar{\nu_e}\to n$ can also be affected by the screening potential, we ignore this reaction %since this three body reaction is not dominant.

\subsection{Effect on the Weak Interactions Rates}
We next consider magnetic field corrections to the weak interaction rates. Electrons and positrons are more sensitive to the background magnetic field than 
the charged baryons because of their smaller masses. 
The thermodynamics of $e^\pm$ will be affected via \emph{Landau quantization}, which has already been addressed in  \citep{Cheng:1993kn,1996PhRvD..54.7207K,1995APh.....3...95G}. In the presence of a magnetic field, the electron (or positron) energy is given by
\begin{equation} \label{eq:1}
E_n^2=p_z^2+m_e^2+2neB,
\end{equation}
After summing over the electron spin, the phase space of electron thermodynamical functions changes to
\begin{equation}\label{eq2}
2\frac{d^3p}{(2\pi)^3}f_{FD}(E;\mu,T) \to \sum_{n=0}^\infty (2-\delta_{n0})\frac{dp_z}{2\pi}\frac{eB}{2\pi} f_{FD}(E_n;\mu,T),
\end{equation}
where the Fermi-Dirac distribution function is one-dimensional.  The
transverse momenta are quantized, resulting in the sum in Eq. (\ref{eq2}). 

\begin{figure}[h!]
\centering
\includegraphics[scale=0.4]{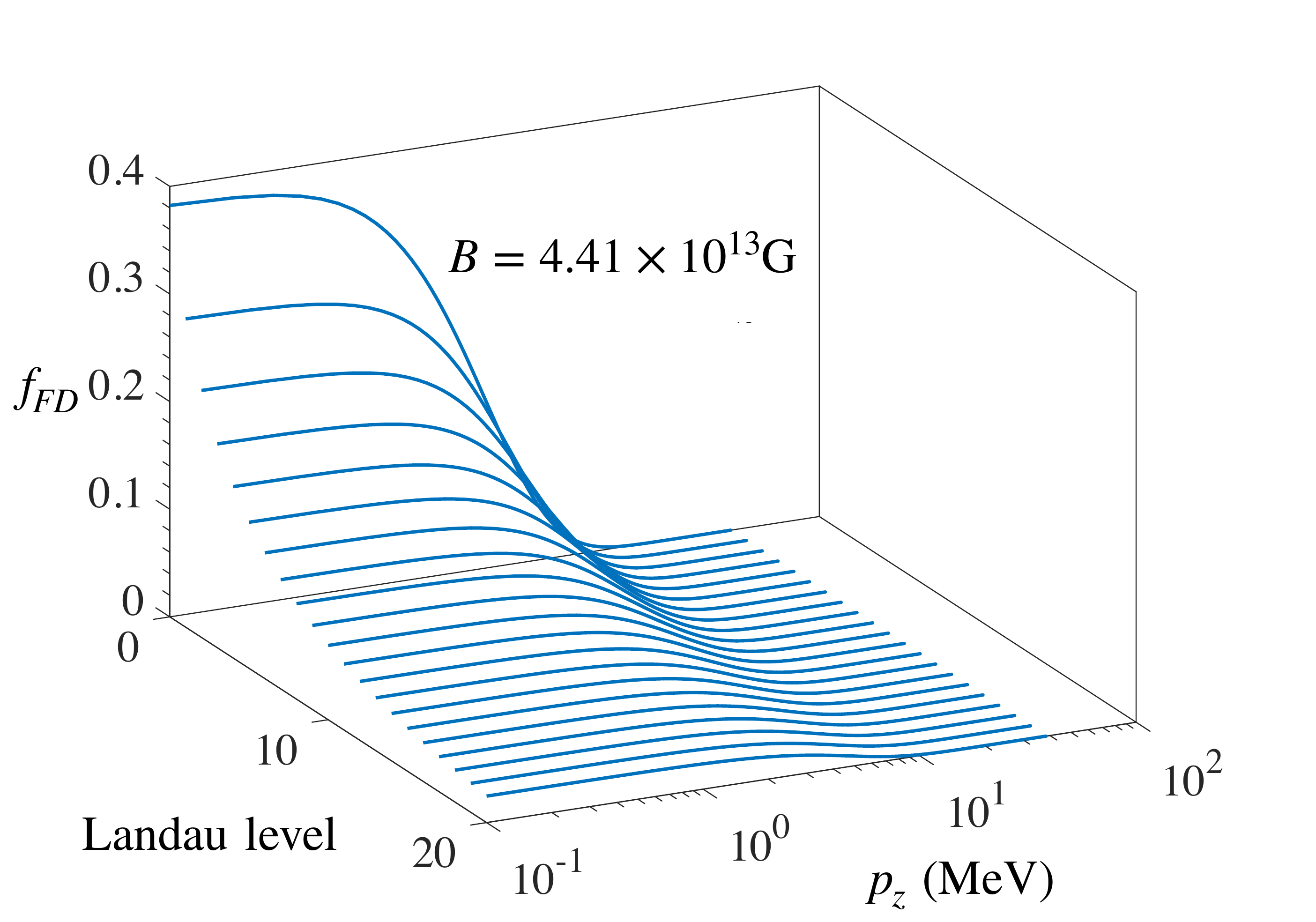}
\includegraphics[scale=0.4]{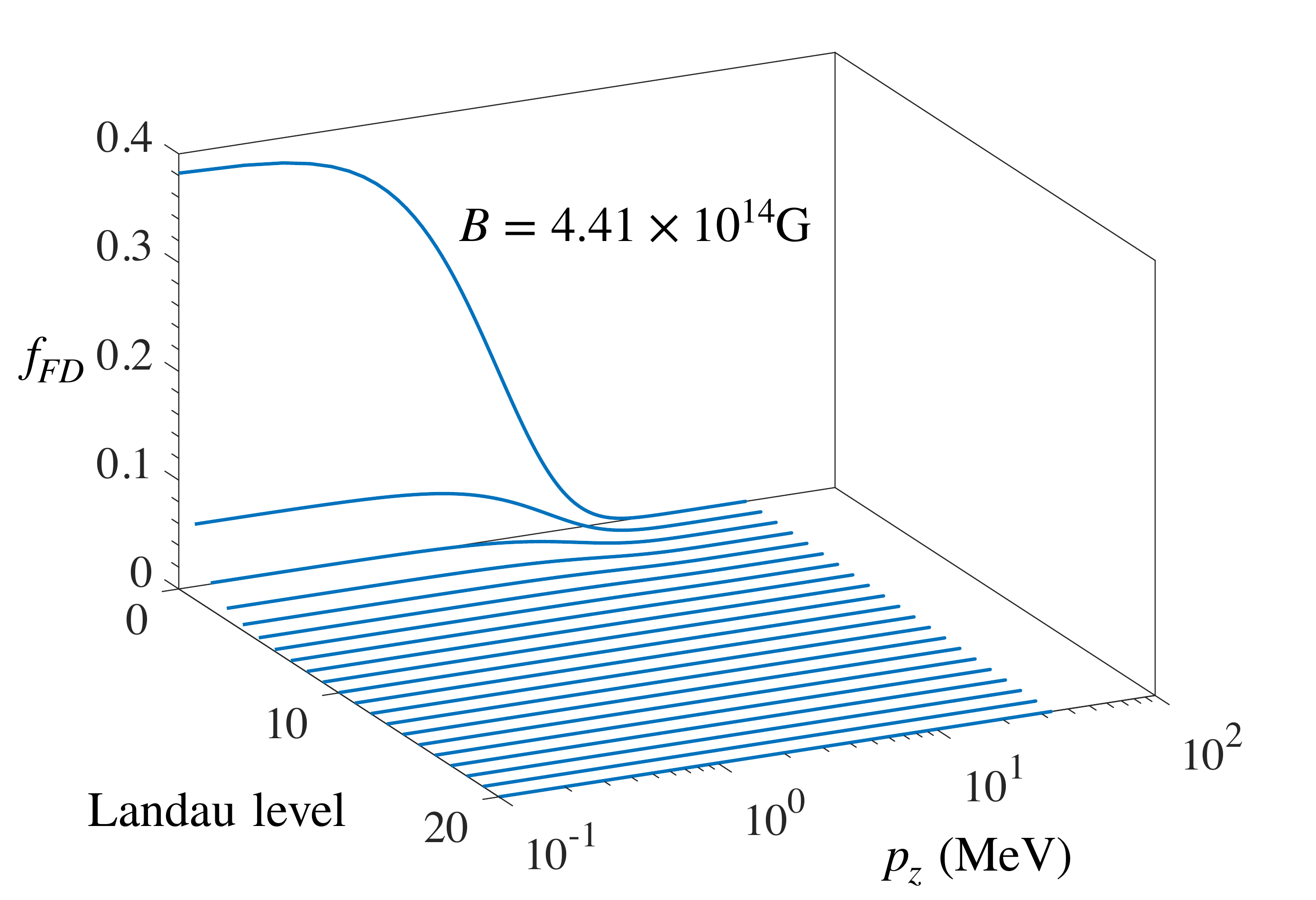}
\caption{\label{fig:3} Fermi distribution functions in the presence of external magnetic fields
as a function of Landua level $n$ and longitudinal momentum $p_z$ for two different field strengths.} 
\end{figure}

In Fig. \ref{fig:3}, we show the distribution as a function $p_z$ and $n$ (i.e. phase space of electrons for each Landau level) for various magnetic
fields. For weak magnetic fields, difference in the  distribution function between two levels is negligible: each distribution approximately equals to the continuous Fermi-Dirac distribution without magnetic fields. For stronger magnetic fields, fermions will occupy  lower Landau levels. It has also 
been pointed out that for strong magnetic field, it is possible to have pair production \cite{1983ApJ...273..761D}, however we here neglect this possibility. 
Including the background magnetic field in the weak interaction rate calculation, Eq. (\ref{eq_debye}) becomes

\begin{align}
\label{debye_l}
\hspace{-10pt}
\frac{1}{\lambda_{TF}^2}=4\pi e^2\frac{\gamma m_e^2}{2\pi^2}\frac{\partial}{\partial\mu}\sum_{n=0}^\infty(2-\delta_{n_0})
\int^\infty_0 dp_z\ \left[\frac{1}{1+\exp{(E_n-\mu)/T}}-\frac{1}{1+\exp{(E_n+\mu)/T}}\right],
\end{align}
where $\gamma$ is the ratio $B/B_c$ with the critical field $B_c$ defined as $B_c\equiv m_e^2/e=4.41\times10^{13}\ \rm G$. 

\begin{figure}[h!]
\centering
\includegraphics[scale=0.45]{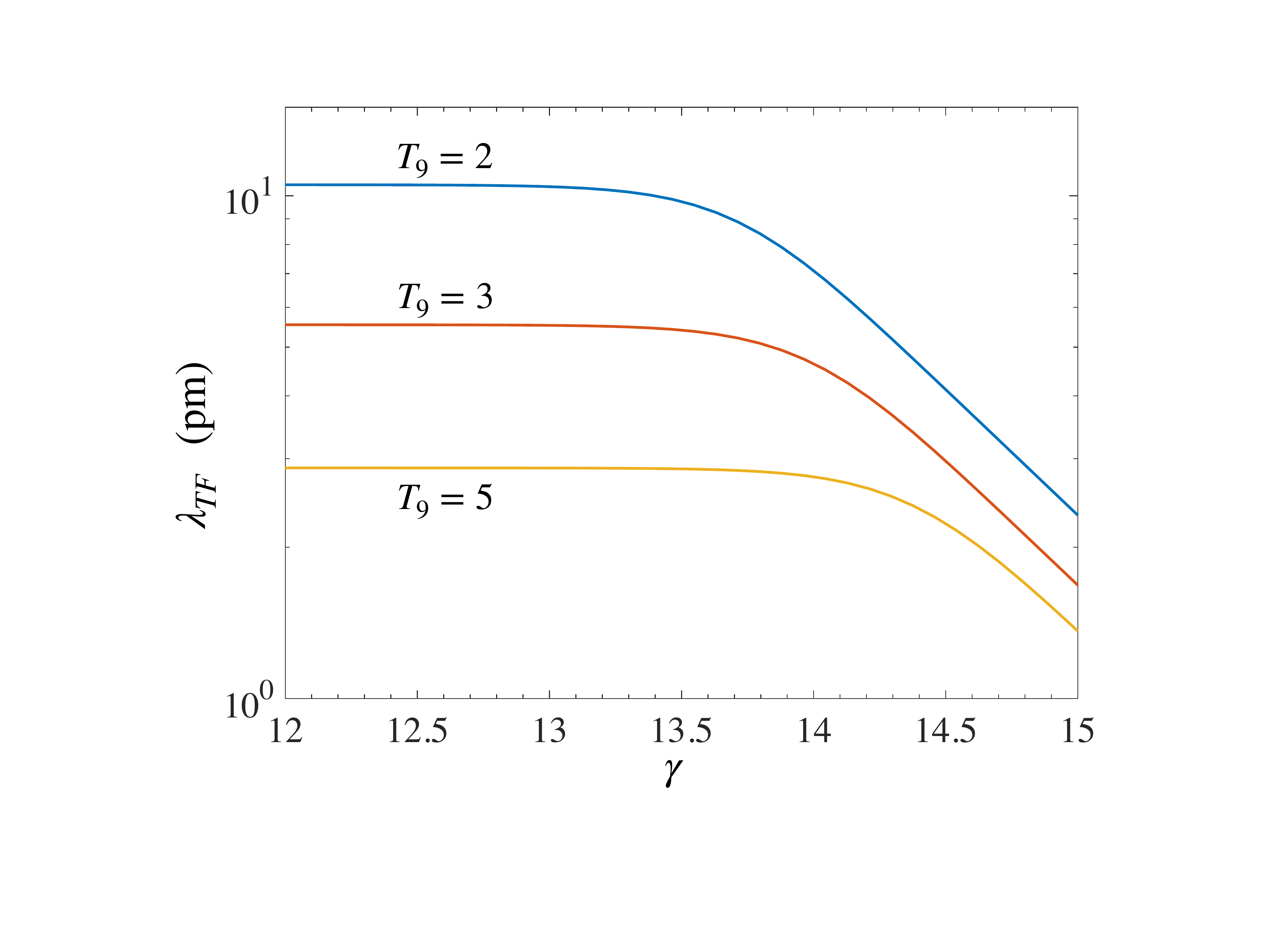}
\caption{\label{fig:1} Thomas-Fermi length, $\lambda_{TF}$, as a function of scaled magnetic field strength, $\gamma$, for different $T_9=T/(10^9\ \rm K)$. The parameter $\gamma$ is defined as $\gamma = eB/m_e^2$. The chemical potential is chosen to be $\mu= 0.1$ MeV.}
\end{figure}

\begin{figure}[h!]
\centering
\includegraphics[scale=0.45]{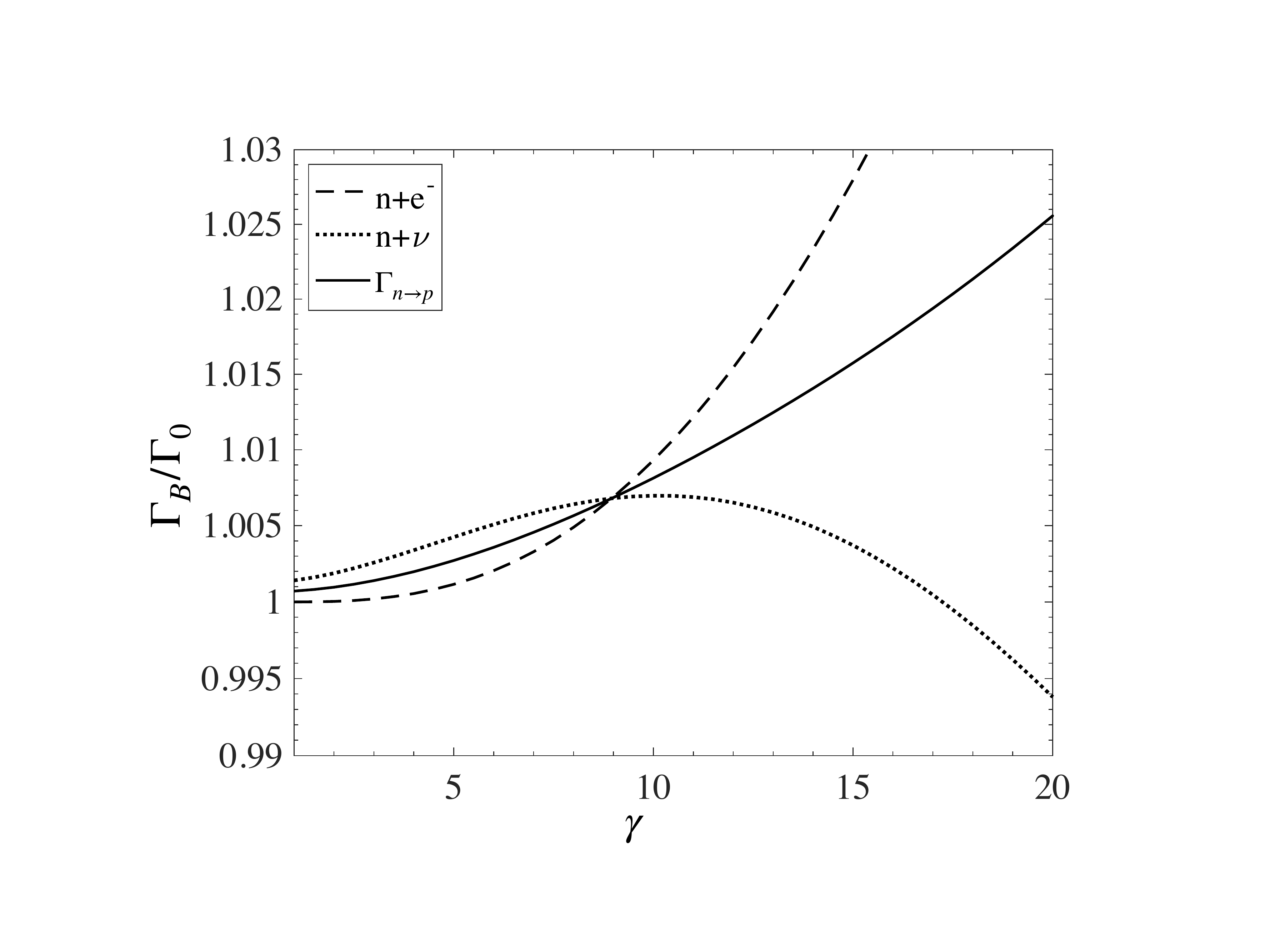}
\caption{\label{fig:weak_rate} Weak interaction rate as a function of the scaled magnetic field strength, $\gamma$. 
The dashed line 
corresponds to the $n+e^+\to p+\bar{\nu_e}$ rate; the dotted line is the $n+\nu_e\to p+e^-$ rate; and the solid line is the total weak interaction rate $\Gamma_{n\rightarrow p}$. For the $n\to p+e^-+\bar{\nu_e}$ rate, we assumed a zero neutrino chemical potential so that this term can be neglected. Here the temperature is set as $T_9=10$.} 
\end{figure}

\begin{figure}[h!]
\centering
\includegraphics[scale=0.4]{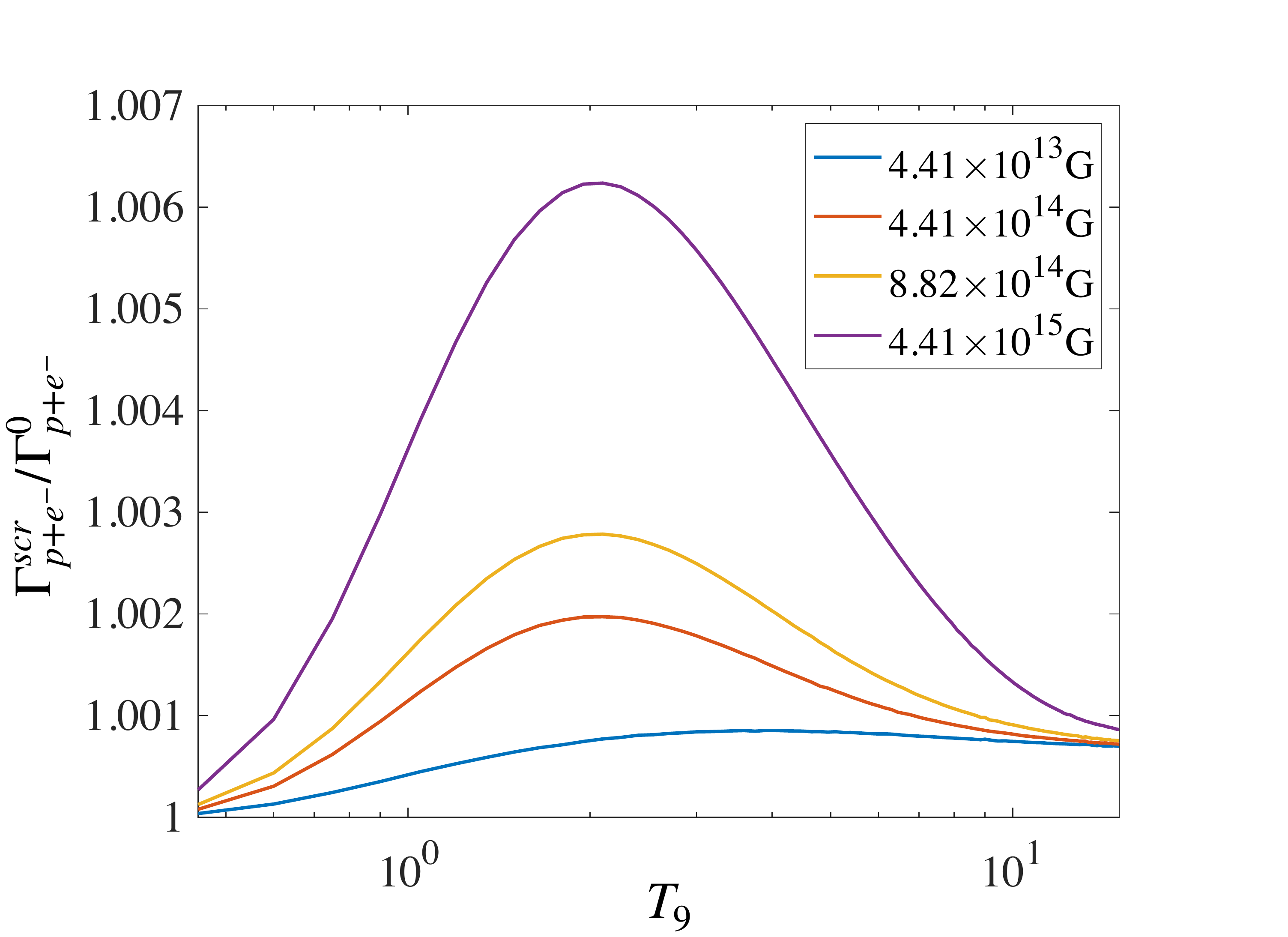}
\includegraphics[scale=0.4]{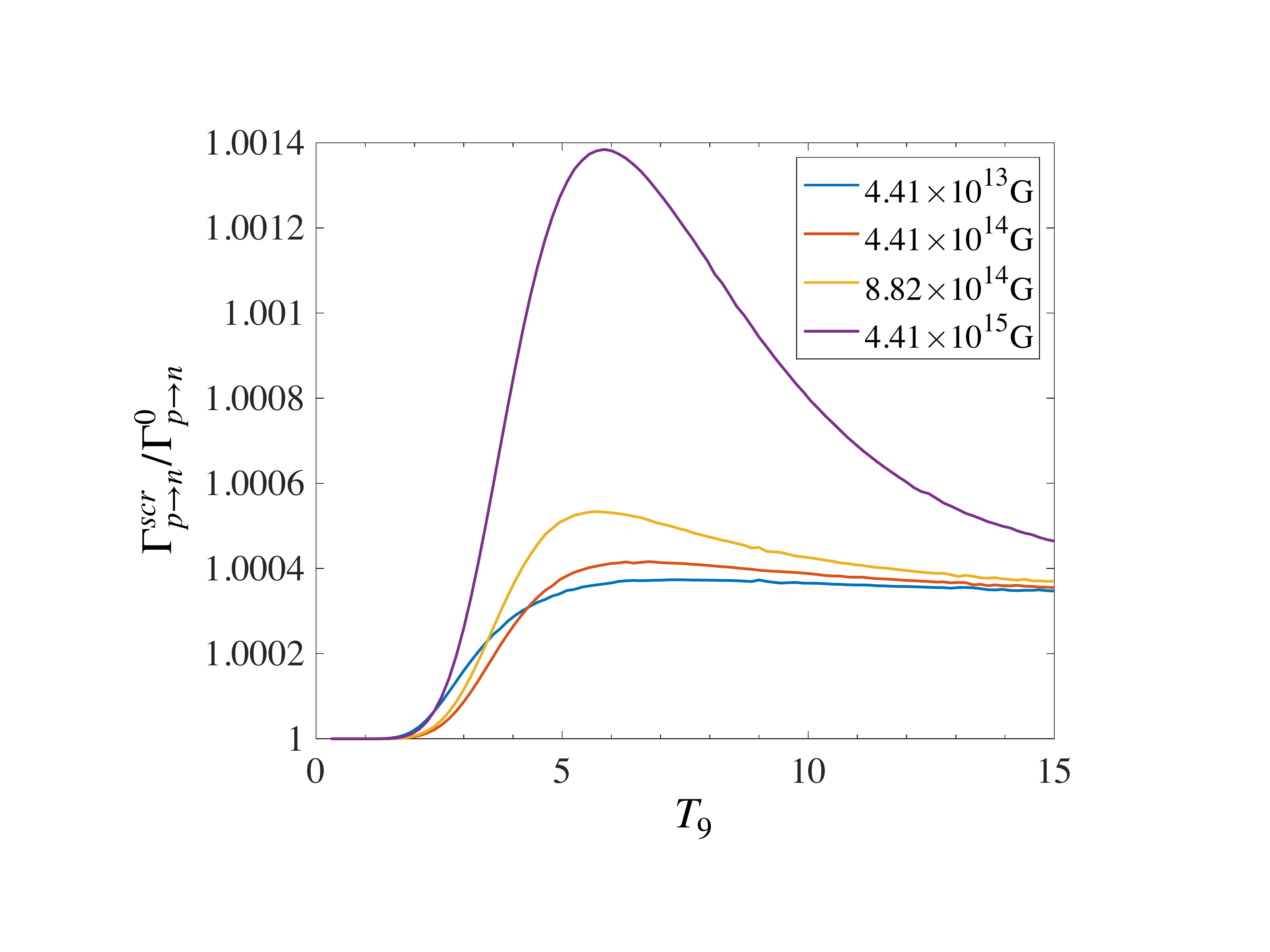}
\caption{\label{fig:5} The screening correction to the $p+e^-\to n+\nu_e$ reaction rate and the total $p\to n$ rate as function of $T_9$ for various field strengths.} 
\end{figure}

Fig. \ref{fig:1} shows $\lambda_{TF}$ as a function of magnetic field strength for
three values of temperature.  In the case of a  weak magnetic field, $B\ll B_c$, $\lambda_{TF}$ does not significantly change.  The change in the distribution functions for different Landau levels is small. For stronger magnetic field strength, $B\gtrsim B_c$, 
$\lambda_{TF}$ drops dramatically. Prior to BBN, i.e. $T\gtrsim1\ \rm MeV$, weak interaction rates  can be strongly dependent on the magnetic field and the temperature. In this epoch $\lambda_{TF}$ is expected to
be much smaller. 
One thus expects an increase of $\Delta V$, altering the electron-capture rate.
With the background magnetic field, it has been
suggested \citep{Cheng:1993kn,1996PhRvD..54.7207K,1995APh.....3...95G} that the weak interaction rate itself also changes due to the Landau quantization. 
There has been some debate as to whether the weak rates increase or decrease as a result of the
magnetic field~\cite{Cheng:1993kn,1996PhRvD..54.7207K,1997PhRvD..56.3766K}.  We show here that the rate of the reaction $n+\nu_e\to p+e^-$ decreases as magnetic field strength increases.  However, the reaction $n+e^+\to p+\bar{\nu_e}$ shows the opposite trend and the summation of two results in a total weak interaction rate $\Gamma_{n\rightarrow p}$ that is enhanced by the existence of
the magnetic field (see Fig. \ref{fig:weak_rate}). Rewriting Eq. (\ref{e_cap_noB}) with the Fermi distribution
given by Eq. (\ref{eq2}), we obtain the electron capture rate 
in a screened plasma:
\begin{eqnarray}\label{correct_B}
\Gamma^{Bscr}_{pe^-\to n\nu_e}=\frac{G_F^2T_\gamma^2(g_V^2+3g_A^2)eB}{\pi^3}\sum_{n=0}^{\infty}(2-\delta_{n0})\nonumber
\\
\times\int_{m_e\sqrt{1+4\gamma n}}^{\infty} \frac{E_\nu^2\epsilon'}{\sqrt{\epsilon'^2-m_e^2(1+4\gamma n)}}\ d\epsilon'\ f_{FD}(\epsilon'; \mu, T_\gamma)g(E_\nu; \mu_\nu, T_\nu).
\end{eqnarray}
In Fig. \ref{fig:5}, we show the weak screening correction of both the electron capture $p+e^-\to n+\nu_e$ and the total $p\to n$ rate as a function of $T_9$. The vertical axis represents the ratio between the interaction rates with and without the screening correction, where the magnetic field effect on the Fermi distribution is included. The weak screening correction increases the electron capture rate (upper panel). Therefore, the total weak reaction rate $\Gamma_{p\to n}$ increases  (lower panel) and finally leads to a higher neutron fraction (see Fig. \ref{fig:corr} below). For a strong B-field (purple line, $\gamma=100$), the impact can be over 0.6\% at $T_9\sim2$. The change itself is small. However, considering the present-day $Y_p$ observation, any corrections which affect weak rates by $\mathcal{O}(10^{-4})$ can be constrained by $Y_p$ abundance observations~\cite{Pitrou2019}, which suggests the possibility of using the weak screening correction to constrain the PMF since the weak interaction plays a leading role before BBN started.

\section{Constraints of Primordial Magnetic field}\label{sect2}
In this section we consider contributions to the final element abundances arising from the weak screening correction of the electron capture rate. Adapting the Thomas-Fermi length formula of Eq. (\ref{eq_debye}), the screening corrections are taken into account using Eq. (\ref{e_cap_noB}). We use a standard BBN nuclear reaction network code \cite{Kawano1992,Smith:1992yy} and have updated the reaction rates of nuclei with mass numbers $A\leq10$ using the JINA REACLIB Database \cite{Cyburt2010,Coc2015}. The neutron lifetime is taken as $880.2 \pm 1.0$ s, corresponding to the central value of the Particle Data Group \cite{Patrignani}. The baryon-to-photon ratio $\eta$ is taken to be $\eta_{10}\equiv\eta/10^{-10}=(6.094 \pm 0.063)$ calculated using a conversion of the baryon density in the standard $\rm \Gamma CDM$ model determined from Planck analysis of \cite{2016A&A...594A..13P}. 
\begin{figure}[h!]
\centering
\includegraphics[scale=0.45]{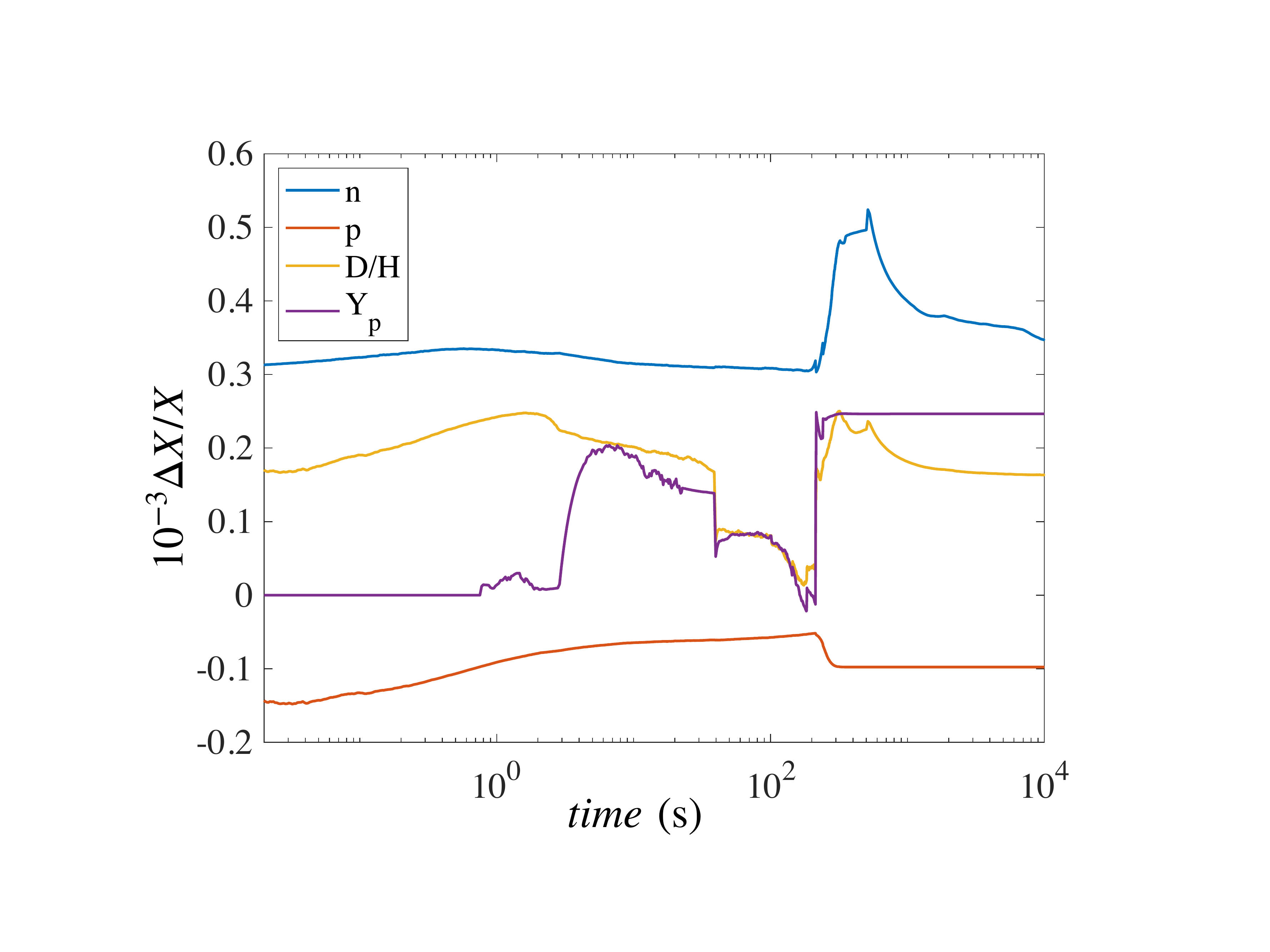}
\caption{\label{fig:corr} Relative change of light nuclear abundances (n, p, D/H and $Y_p$) due to the weak screening correction on the electron capture reaction, i.e. $p+e^-\to n+\nu_e$, in the BBN network as a function of time. Effects from the magnetic field on the Fermi distribution function at the relevant BBN temperatures are negligible as shown in Figure \ref{fig:1}.} 
\end{figure}

Fig. \ref{fig:corr} shows the ratios of final abundances of light nuclei (n, p, D/H and $Y_p$) with weak screening effects on the electron capture rate to those calculated without screening effects. The quantity $Y_p$ is effectively determined by $2n/(n+p)$ at the $^4$He synthesis at $t\sim180\ \rm s$. Therefore, the higher neutron fraction naturally leads to a higher $^4$He mass fraction. 

We consider a constraint on generation epoch of the PMF. In this study, we employ the "frozen-in" PMF model, i.e., the PMF energy density decreases as $\rho_{PMF}\propto 1/a^4$ where $a$ is the scale factor of the Universe. Current constraints on the PMF from light element abundance observations can only provide us with an upper limit of the field strength \cite{1995APh.....3...95G,1996PhRvD..54.7207K,Kawasaki:2012kn}. 
We have investigated three main effects from the PMF on the electron/positron thermodynamics, the time-temperature relation, and thermonuclear reaction rates \cite{Kawasaki:2012kn}. The impact on the weak interaction rates are always neglected due to the large uncertainty of past $Y_p$ observations~\cite{Kawasaki:2012kn}. However, the updated observational constraint on primordial $^4$He abundance is accurate to within $0.1\%$.

\begin{figure}[h]
\centering
\includegraphics[scale=0.4]{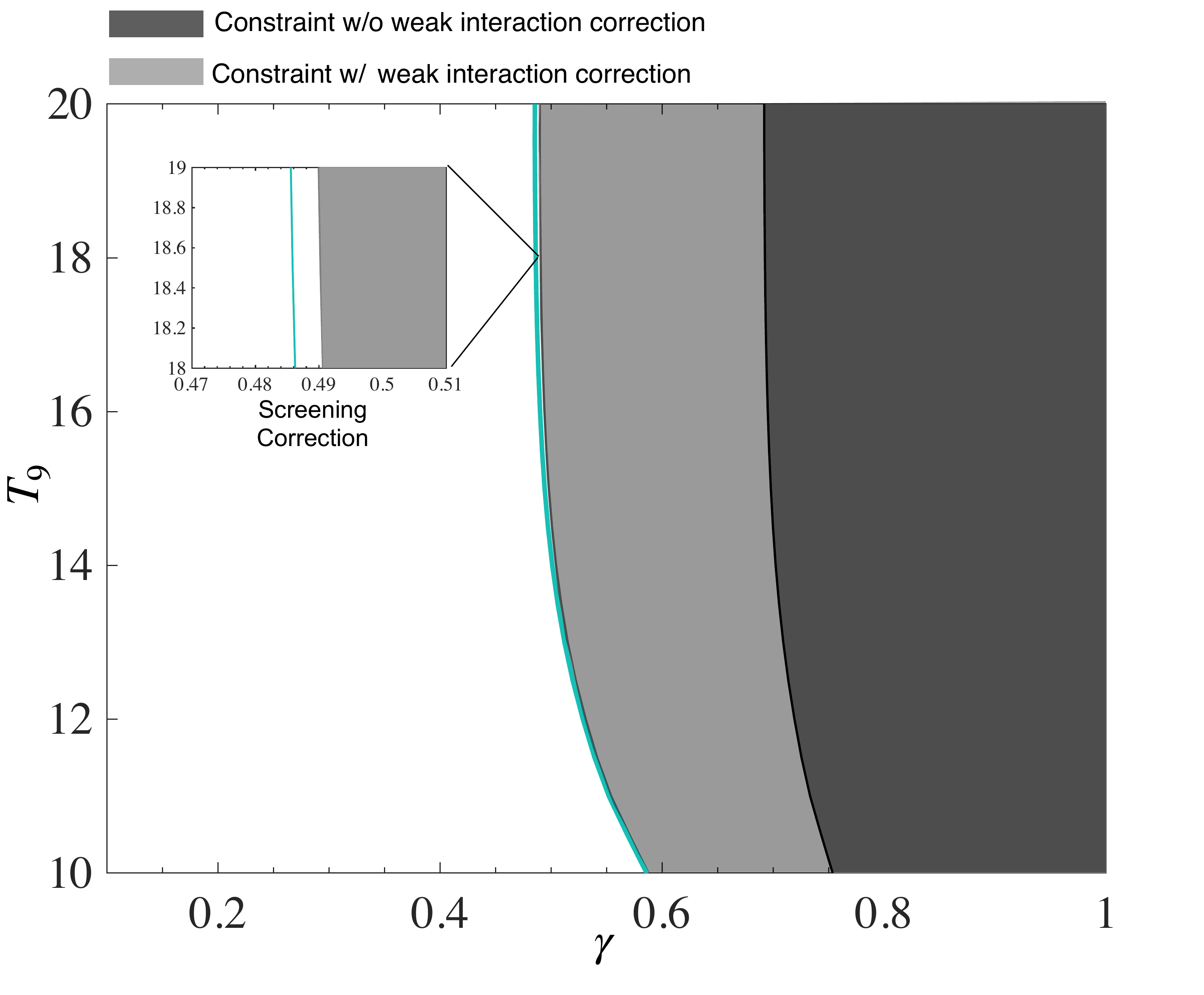}
\caption{\label{T_B} Constraints of the PMF generation epoch and strength from the $Y_p$ observational value. The light gray shaded region is excluded if the modification of weak reaction rates by the magnetic field is taken into account.  The dark gray region is excluded by prior work~\cite{Yamazaki2012,Kawasaki:2012kn}, in their study the PMF impacts on weak interaction are ignored. The constraint from the screening correction of weak reaction rates is shown by the blue line.  This constraint is negligible since the density of  electrons and positrons during the BBN epoch is low. Here the $\gamma$ value of the PMF is taken at $T_9=10$.} 
\end{figure}

Fig. \ref{T_B} shows the constraint on the generation epoch and the strength of the PMF. The horizontal axis is the
strength of the PMF in unit of $B_c$ at $T_9=10$, and the vertical axis is the temperature at which the PMF is generated.
We only consider the PMF generated before the neutrino decoupling at $T\sim1\ \rm MeV$, and vertical axis is only shown
above $T_9=10$ accordingly. We encoded the "frozen-in" PMF generated at different temperatures and then performed the
BBN calculations. The shaded region on the right-upper part of the figures is ruled out by $Y_p$ observations $Y_p =
(0.2449\pm0.0040)$ \cite{Aver}. Although the $^4$He abundance is sensitive to the n/p ratio, for the lately ($T_9<15$)
generated PMF the constraint is weaker since the weak reaction rates drop quickly when temperature decreases. Thus, such
a PMF cannot alter $Y_p$ as significantly as the early generated PMF, which means one can introduce a stronger PMF at
later times without changing the calculated $^4$He abundance. The enhancement of weak interaction rates induces a tighter constraint on the PMF. The weak screening correction to fusion reactions does not make a significant change in BBN due to the low electron-positron density at the BBN epoch.

%We note that the PMF also affects the weak reaction rates by changing the masses of neutrons and protons. However, this
%effect is negligibly small since it requires a magnetic field as strong as $\mathcal{O}(10^{18})$ G to significantly 
%change the nucleon masses \cite{Bander:1992ku}. 
In Fig. \ref{T_B}, all effects from the PMF summarized in Ref. \cite{Kawasaki:2012kn} have been taken into account (shown in the dark gray region). A more accurate constraint on the B 
field based on the consideration of the weak interaction rate enhancements via the PMF is shown by the light gray region.
It is clearly seen that such effects can provide a narrower constraint on the PMF strength. Because weak interactions
decouple at $T\sim 0.8\ \rm MeV$, the PMF generated well before this epoch plays an important role in determining the
light element abundances. According to Fig. \ref{fig:5}, the screening corrections can increased with increasing magnetic
field. This is also taken into account and indicated by a blue line.

\begin{figure}[h]
\centering
\includegraphics[scale=0.4]{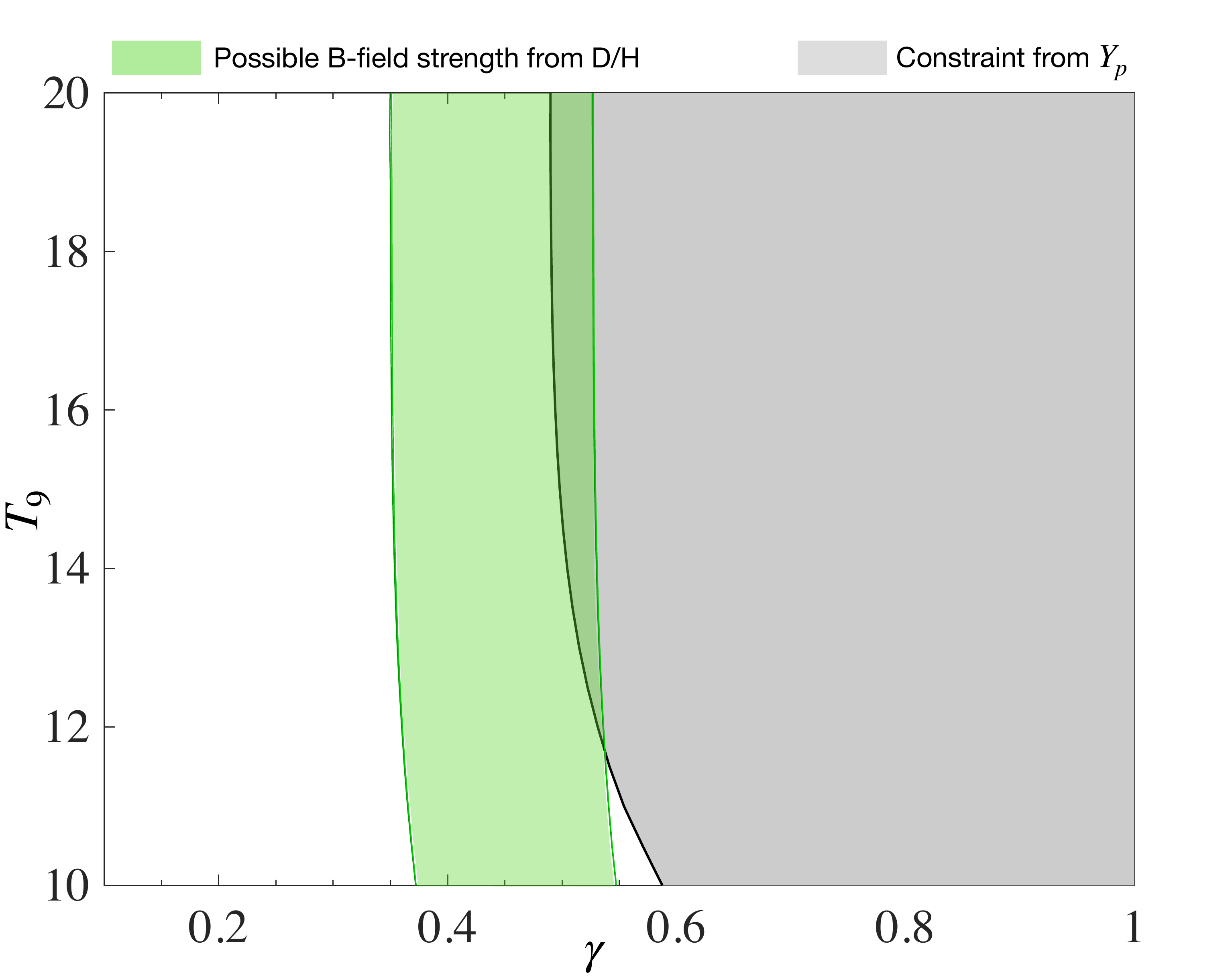}
\caption{\label{D_He} Range of PMF strength constrained by deuterium abundance observations, i.e., D/H $=(2.527\pm 0.03) \times 10^{-5}$. Here the scaled value of the PMF, $\gamma$, is taken at $T_9=10$. The gray region is excluded by constraints from
$^4$He abundance observations while the green
region is allowed by deuterium abundance observations.} 
\end{figure}

Recent high-accuracy BBN calculations suggest an underproduction of D for $\eta_{10}=6.10$ when compared to the mean value of the D observation, i.e., D/H$=(2.527\pm 0.03)\times 10^{-5}$ \cite{Cooke2018}. {Uncertainties in nuclear reaction rates for D destruction result in a $\sim 1.5$ \% error in the predicted D abundance, i.e., D/H=$(2.459 \pm 0.036)\times 10^{-5}$ \cite{Pitrou2019}. Therefore, there is a possible discrepancy at  $\sim 2\sigma$ level.} We also consider the solution of such discrepancy from the standpoint of modifications of weak and fusion reactions by the PMF. We have already shown that the $^4$He abundance constraint allows a PMF with $\gamma<0.58$. Moreover, when we also take the D/H constraint into account,  recent observations can actually exhibit clear discrepancy with PMF since both D/H and $^4$He abundances are enhanced when PMF is included \cite{1995APh.....3...95G,Kawasaki:2012kn,Yamazaki2012}. In Fig. \ref{D_He}, we show the contour plot of both D/H and $Y_p$ observational abundance constraints. The green region is the observational constraint D/H $=(2.527\pm 0.03) \times 10^{-5}$, and it is clear that for the PMF model with strength parameter $\gamma=0.37-0.54$, the D/H prediction is consistent with the observation. Such a PMF is not ruled out by taking account of the $Y_p$ observational constraint as well. {\color{blue}If the "D underproduction problem" were confirmed in extensive and more accurate observations in the future, it would be provide an additional explanation for faster cosmic expansion triggered by a large effective number of neutrino families $N_{\rm eff}$ \cite{Cyburt:2016cr} due to a sterile neutrino or something equivalent to it.}

 { In Table \ref{tab1}, we compare the observational constraints on primordial abundances with the theoretical predictions in three models, i.e., (1) the SBBN, (2) the BBN model with the screening correction, and (3) the BBN model with the screening correction and PMF effects for $\gamma=0.4$, for example. Although the "D underproduction problem" in the SBBN is not solved in model (2) because of its very small effect, it is solved when we introduce a "frozen-in" PMF with strength $\gamma=0.37-0.54$ in the model (3).  Neither the screening effect nor the "frozen-in" PMF model can alleviate the cosmic lithium problem.  Uncertainties in the nuclear reaction rates for $^7$Be production and destruction have been reduced by recent experiments on $^7$Be destruction reactions. As for the primary destruction reaction $^7$Be($n$,$p$)$^7$Li, a recent measurement at the neutron time-of-flight (n\_TOF) facility of CERN showed that the cross section is significantly higher than previous measurements in the low neutron energy region of $E_n \sim10^{-2}$ MeV, while it is consistent with the old measurements for higher energies \cite{Damone:2018mcf}. 
 
 Replacing 
 the old reaction rate by the newly derived rate, the predicted $^7$Li abundance becomes smaller by at most 12 \%. The effect of including the first excited state of $^7$Li in the final state is now under analysis utilizing $Q$-value spectra of the $^7$Be(d,$^7$Li\ p)$^1$H and $^7$Be(d,2$\alpha$)$^1$H reactions which are generated with the Center for Nuclear Study Radioactive Isotope Beam separator \cite{Hayakawa2019}. The contribution of  the first excited state is estimated to be at most $\sim 15$ \%, further reducing the predicted $^7$Li abundance. The reaction cross section of $^7$Be(d,2$\alpha$)$^1$H has been recently measured in the energy range relevant to BBN \cite{Rijal:2018vbz}. The new cross section leads to a $1.4$\%--$8.1$\% decrease of the primordial $^7$Li abundance compared to the case without the $^7$Be(d,2$\alpha$)$^1$H reaction. The cross section of the reaction $^7$Be(n,$\alpha$)$^4$He has also been determined precisely at the n\_TOF facility in CERN ($E_n \lesssim 10$ keV) \cite{Barbagallo:2016zxk}, the n\_TOF in the Research Center for Nuclear Physics, Osaka University (the center of mass energy $E_{\rm CM} =0.20$--0.81 MeV) \cite{Kawabata2017}, and the EXOTIC facility of Laboratori Nazionali di Legnaro ($E_n =0.03$--2 MeV) \cite{Lamia2019}. The contribution of this reaction to the destruction of $^7$Be during the BBN was found to be negligibly small.} 

\begin{table}%[H] add [H] placement to break table across pages
  \caption{\label{tab1} Comparison between observations and theoretical predictions for primordial abundances. Here in the theoretical calculation, all the cross sections for nuclei $A<10$ are adopted from the JINA REACLIB Database \cite{Cyburt2010,Coc2015}. The neutron lifetime is taken as $880.2$ s \cite{Patrignani}, the baryon-to-photon ratio $\eta$ is taken to be $\eta_{10}\equiv\eta/10^{-10}=(6.094 \pm 0.063)$ \cite{2016A&A...594A..13P}. We use the PMF model with strength parameter $\gamma=0.4$.}
      %References to adopted cross sections ---Already done (Yudong)
   \centering  
    \begin{tabular}{c c c c c}
    \hline\hline
       & &        & BBN &  BBN \\ [-2.8 ex]
        &Observ. & SBBN& + Screening Corr.& + Screening Corr.\\[-2.2 ex]
       & & & & +PMF\\ [+0.5 ex]
      \hline
 $Y_p$        &      $0.2449\pm0.0040 $(a)                        & $0.2417\pm0.0001$  &$0.24165\pm0.00005$  &$0.2477\pm0.0001$  \\
 D/H $(\times10^{-5})$  &    $2.527\pm0.03$(b)                                  &   $2.462\pm0.042$   & $2.462\pm0.042$ &$2.545\pm0.043$  \\
$A=7$ $(\times10^{-10})$ & $1.58^{+0.35}_{-0.28}$(c)                               &  $4.90\pm0.105$    &$4.90\pm0.105$  &$4.87\pm0.11$  \\
\hline\hline
    \end{tabular}
(a)Aver {\it{et al.}} (2015) \cite{Aver},(b)Cooke {\it{et al.}} (2018) \cite{Cooke2018},(c)Sbordone {\it{et al.}} (2010) \cite{2010A&A...522A..26S}.
\end{table}

\section{Conclusions}\label{sect3}
In this paper, we investigated weak screening corrections from the PMF and the impact of these corrections on the electron capture rate.  The lowest-order screening effect is to shift kinetic energies of electrons and positrons. We first explored the impact from such corrections on the BBN, and found an  enhancement of the $^4$He abundance by a factor of $\mathcal{O}(10^{-4})$. Then, we considered the configuration with a  background PMF in which the electron and positron energy distributions are altered by Landau quantization. The presence of an external magnetic field results in a shift in the screening potential. Moreover, with the existence of an external magnetic field, the weak screening correction can enhance the electron capture rate by a factor of $\mathcal{O}(10^{-3})$.  Such effects on the electron capture rate can be negligible due to the low density at BBN epoch.

We show that the magnetic field results in a reduction of the rate for the reaction $n+\nu_e\to p+e^-$ while the rate
for the $n+e^+\to p+\bar{\nu_e}$ reaction is increased. The net rate $\Gamma_{n\rightarrow p}$ turns out to be enhanced by the magnetic field effects. We conclude that such an enhancement of weak reaction rates from the background PMF should be taken into account since the accuracy of present-day theoretical calculations requires detailed treatments of any change of weak reaction rate larger than $0.1\%$. 

{Finally, the generation epoch of a "frozen-in" PMF has been constrained by considering its impact on weak interactions. Comparing the theoretical $^4$He yield with  observations, we find that a late PMF generation epoch at $T_9<15$ is more favored.

Moreover, the ``D underproduction  problem'' in SBBN could be solved by including the effects of the PMF, resulting in an enhancement of weak reaction rates. Namely, we find an allowed region which satisfies both of D/H and $Y_p$ observational abundance constraints in the BBN model with the screening correction and PMF effects. 

However, the cosmic lithium problem still remains. Possible solutions to this problem include the following scenarios: (1) BBN models with exotic long-lived negatively-charged particles \cite{Jittoh:2007fr,Bird:2007ge} or a color \cite{Kawasaki:2010yh} can potentially solve the problem. (2) The existence of a sterile neutrino during the BBN can reduce the $^7$Li abundance significantly only if its mass and lifetime are in specific ranges \cite{Kusakabe:2013sna,Ishida:2014wqa}. 
(3) An ambipolar diffusion of abundant $^7$Li$^+$ ions via the PMF during structure formation can result in Li abundances in structures smaller than the cosmic average value \cite{Kusakabe:2014dta,Kusakabe2019}.
(4) If population III (Pop III) stars deplete Li with a very large formation rate and if they do not produce Li via the neutrino process, the Li abundance can temporarily decrease in the early structure formation epoch \cite{Piau:2006sw}. However, a recent calculation of the neutrino process in Pop III stars indicates efficient Li production \cite{Heger:2020jbp}. In this case, the Li abundance monotonically increases with time, and this scenario does not provide a solution.  However, there remain possibilities of significant Li depletion. (5) $^7$Li could be destroyed in a highly convective pre-main sequence stage via nuclear burning \cite{Fu2015} and also (6) during the main sequence via atomic diffusion under stellar gravity \cite{Korn:2006tv}.

}

\begin{acknowledgments}
Y.L. is supported by JSPS KAKENHI Grant No. 19J22167; M.A.F. is supported in part by NASA Grant No. 
80NSSC20K0498; T.K. is
supported in part by Grants-in-Aid for Scientific Research of JSPS (15H03665 and 17K05457); M.K. is supported 
by NSFC Research Fund for International Young Scientists (11850410441). A.B.B. is supported in part by the U.S. 
National Science Foundation
Grant No. PHY-1806368.  M.A.F., M.K., and A.B.B. acknowledge support from the NAOJ Visiting Professor program.

\end{acknowledgments}

\end{document}